\documentstyle[preprint,aps,epsf]{revtex}

\begin{document}

\title{Biscale Chaos in Propagating Fronts}
\author{Anatoly Malevanets, Agust\'\i\ Careta$^*$ and Raymond Kapral\\
Chemical Physics Theory Group, Department of
Chemistry, \\University of Toronto, Toronto, Canada M5S 1A1}
%\date{\today}

\maketitle

\begin{abstract}
The propagating chemical fronts found in cubic autocatalytic
reaction-diffusion processes are studied. Simulations of the 
reaction-diffusion equation near to and far from the
onset of the front instability are performed and the structure and 
dynamics of chemical fronts are studied. Qualitatively different front
dynamics are observed in these two regimes. Close to onset the front 
dynamics can be characterized by a single length scale and described
by the Kuramoto-Sivashinsky equation. Far from onset the front
dynamics exhibits two characteristic lengths and cannot be modeled 
by this amplitude equation. An amplitude equation is proposed for
this biscale chaos. The reduction of the cubic autocatalysis
reaction-diffusion equation to the Kuramoto-Sivashinsky equation is
explicitly carried out. The critical diffusion ratio $\delta_c$, where the 
planar front loses its stability to transverse perturbations, is
determined and found to be $\delta_c=2.300$.
\end{abstract}

\section{Introduction}
Propagating fronts separating regions with different characteristics 
occur in many physical contexts. Often such fronts have a complicated 
structure of their own as in the fractal forms that arise in some 
diffusion-limited aggregration processes or viscous 
fingering\cite{stan:gro}. In this article we consider
chemical fronts separating regions of distinctly different chemical
concentrations and study the nature of the spatio-temporal dynamics that
these fronts exhibit. In these systems, under appropriate conditions, a
planar front may not be stable and complex, even chaotic, front dynamics 
can arise. For example, this is the situation often encountered in 
propagating flame fronts which have been studied extensively\cite{flames}. 
The complex dynamics of fronts can also
underlie and determine the character of spatio-temporal chaos and dynamics
in a variety of other chemical contexts\cite{meron}.

We study a specific system here, the fronts in the cubic autocatalysis
reaction $A + 2B \rightarrow 3B$, but several parts of our analysis are
general and should apply in other situations. These cubic autocatalysis
fronts have many features in common with propagating flames\cite{siva:inst}. 
An experimental realization of such chemical front instabilities occurs in the 
iodate-arsenous acid reaction carried out in a gelled medium to 
suppress fluid flow\cite{horv:ins2}. The experiments show patterned 
fronts, much like those in quadratic and cubic mixed-order models.

For the cubic autocatalysis model, when the diffusion coefficient of the fuel 
$A$ is sufficiently larger than that of the autocatalyst $B$, 
the planar front is unstable to transverse perturbations. We have investigated 
the nature of the resulting front dynamics as a function of the diffusion 
ratio $D_A/D_B$ in large systems. Several results have emerged from our 
study of this system. For small enough diffusion
ratios within the unstable regime the front exhibits complicated 
dynamics with statistically
stationary properties and is characterized by a single length scale. 
Far beyond the instability point a new chaotic regime is encountered, 
termed biscale chaos, in which there are two characteristic length scales. 

We have carried out a detailed analysis of the front dynamics in terms of
amplitude equations. Just beyond the instability 
threshold the front dynamics is described in terms of the
Kuramoto-Sivashinsky equation\cite{kur:orig,siv:orig}. 
We document this relation through a
statistical characterization of the interface. The regime far from the 
instability point where biscale chaos is observed cannot be described in 
terms of the
Kuramoto-Sivashinsky equation and we construct an amplitude equation in
which the nonlinear mode coupling is generalized. This amplitude equation
is able to capture the principal qualitative features of the biscale chaos. This 
part of the analysis may generalize to other situations. 

We also give a detailed reduction of the cubic autocatalysis
reaction-diffusion equation to the Kuramoto-Sivashinsky equation. The
coefficients that enter in this equation are estimated analytically using
solutions for the right and left eigenvectors of the eigenvalue problem
that enters the analysis, and a numerical scheme is devised that allows the
computation of these coefficients for any values of the diffusion
coefficients. As a by-product of this analysis we may easily determine the
critical diffusion ratio where the planar front becomes unstable.  

\section{Cubic Autocatalysis Fronts} \label{sec:phen}
Consider the autocatalytic reaction $A+2B \longrightarrow 3B$ described by 
the reaction-diffusion equation,
\begin{eqnarray}
{\partial \alpha \over \partial t}&=& -\alpha \beta^2 +D_A \Delta 
\alpha \;, \nonumber \\
{\partial \beta \over \partial t}&=& \alpha \beta^2 + D_B \Delta
\beta\;,
\label{rdeq}
\end{eqnarray}
where $\alpha(t,{\bf r})$ and $\beta(t,{\bf r})$ are the (scaled) 
concentrations of the $A$ and
$B$ species, respectively, with diffusion coefficients $D_A$ and
$D_B$. In one dimension, with suitably defined initial
conditions\cite{bill:deve}, this system supports isothermal chemical 
fronts\cite{scot:simp}.

In two space dimensions the planar front is unstable to transverse 
perturbations when $D_A$ is sufficiently larger than $D_B$. The origin 
of such instabilities and the nature of the resulting front dynamics 
have been the subject of earlier studies\cite{siva:inst}. 
Horvath et al.\cite{horv:inst}
performed simulations of the cubic autocatalysis model (\ref{rdeq}) and
investigated the bifurcation structure as a function of the system length. 
For small system sizes, when the front possesses a single minimum, the 
onset of chaotic behaviour was observed to occur through a period 
doubling cascade in the temporal dynamics of the minimum. 
For large system sizes the 
resulting front dynamics is best analysed in statistical terms. We
now briefly describe and quantitatively characterize 
the nature of the front dynamics for large system lengths as a function 
of the diffusion coefficient ratio, $\delta=D_A/D_B$.

We consider a two-dimensional system which is infinite in the $x$ direction 
and has length $L$ along $y$, where periodic boundary conditions are applied. 
The initial conditions are
taken to be $\{\alpha(0,x,y)=1,\;\beta(0,x,y)=0\; |\; x \le -w/2, \forall y\}$ and 
$\{\alpha(0,x,y)=0,\;\beta(0,x,y)=1\; |\; x \ge w/2, \forall y\}$. The 
region $\{x,y\;|\;-w/2 < x < w/2, 0 \le y \le L\}$ was divided into segments 
of length $\ell$ along $y$ and each
segment was assigned the values $(\alpha(0,x,y),\beta(0,x,y))=(1,0)$ or 
$(0,1)$ with probability $p=1/2$. It is
convenient to represent the front dynamics in terms of the
isoconcentration profiles, $h_A(t,y)$ and $h_B(t,y)$ defined by 
$\alpha(t,h_A(t,y),y)=\alpha_r$ and $\beta(t,h_B(t,y),y)=\beta_r$, where
$\alpha_r$ and $\beta_r$ are reference concentrations. Henceforth we focus
on $h_B(t,y)$ measured relative to its average value at time $t$ 
and denote this quantity by $h(t,y)$. We note that for these initial 
conditions this average value moves with the minimum front speed, which 
is selected from the continuum of allowed front speed values.

Figure~\ref{fig:min}(a) shows a space-time plot of the minima of 
$h(t,y)$ for $\delta=5$. In the integration\cite{euler} of (\ref{rdeq}) we 
have scaled the length variables so that $D_A=1$ and $D_B=\delta^{-1}$. 
After an initial transient period during which the 
interface develops, a
moving front with statistically stationary properties is formed. In this
stationary regime the spatio-temporal behaviour can be described in terms of
the dynamics of the minima which play the role of ``particles" in the 
system. The particles collide and coalesce and new particles are born so that the
average density of particles per unit length remains constant. The
front dynamics may be characterized quantitatively by the space-time 
correlation function
\begin{equation}
C(t,y)= \langle L^{-1} \int_0^L dy' [h(t,y') - h(t,y+y')]^2 \rangle\;,
\label{correq}
\end{equation} 
where the angle brackets signify an average over realizations of the
front evolution starting from the random initial conditions given above. 
We also consider the time average of $C(t,y)$ in the stationary regime,
\begin{equation}
\bar{C}(y) = T^{-1} \int_{t_0}^{t_0+T} dt'\; C(t',y)\;,
\label{spacecor}
\end{equation} 
where $t_0$ is a time longer than the transient period. 
The correlation function (\ref{correq}) is sketched in
Fig.~\ref{fig:corplot}
for several values of the time $t$ and shows the development of
correlations during the transient period and the approach to the
stationary regime. The function $C(t,y)$ exhibits short range 
correlations with a tendency to reach a nearly
parabolic form, seen in the $\bar{C}(y)$ curve, which is characteristic 
of diffusive transverse front dynamics and will be discussed further below. The
persistent short range correlations are due to the existence of a
characteristic length $\ell^*$ arising from the front instability.

This type of behaviour persists up to about $\delta=6$ when a new type of
chaotic front dynamics, distinguished by the appearance of a second
length scale, is observed. The second length scale 
is easily discerned in both the space-time plot of the minima of $h(t,y)$
shown in Fig.~\ref{fig:min}(b) for $\delta=8$ and the power spectrum,
\begin{equation}
E(k)= T^{-1} \int_{t_0}^{t_0+T} dt \langle |h(t,k)|^2 \rangle \;.
\end{equation}
The power spectrum is presented in Fig.~\ref{fig:pspec} for both $\delta=5$ 
and $\delta=8$. For $\delta=5$ one sees a single peak corresponding to the 
wavenumber of the most unstable mode. However, for $\delta=8$ the minimum 
in the power spectrum has filled in, indicating the growth of modes 
with smaller wavenumbers and thus the appearance of structure 
on longer length scales. We now turn to an analysis of these results in
terms of amplitude equations and provide a foundation for the phenomenological 
picture of the front dynamics.

\section{Analysis of Front Dynamics} \label{sec:anal}
A description of the origin of the front dynamics is
most conveniently given in terms of amplitude equations for the front
profile derived from the reaction-diffusion equation (\ref{rdeq}). We
analyse the dynamics close to the instability point in terms of
the Kuramoto-Sivashinsky equation and show how this equation must be
generalized in order to explain the regime far beyond the instability 
where a second characteristic length appears.

\subsection{Dynamics of small perturbations}
We present an adaptation of Kuramoto's 
derivation\cite{kura:chem} for the case of small amplitude perturbations 
which yields the amplitude equation that will be used in the
subsequent analyses. 
We shall confine our attention to the question of how the presence
of a second dimension affects the dynamics of the general reaction-diffusion 
system,
\begin{equation}
\label{reaction}
{\partial {\bf z} \over \partial t}= {\bf F(\bf z)} + {\bf D}\Delta{\bf z}\;.
\end{equation}
Here ${\bf z}$ is a vector of concentration fields and ${\bf F(\bf z)}$ is
a vector-valued function describing chemical reactions. 
We assume that in one dimension (\ref{reaction})
possesses a stable solution with a  propagating front profile
$ {\bf z}(t,x,y) = {\bf z}_0(x-ct) $, where $c$ is the
velocity of the front. Following the treatment by Kuramoto\cite{kura:chem}
we seek the dynamics of a perturbed solution in the form,
\begin{equation}
\label{peturbed}
	{\bf z}(t,x,y) = {\bf z}_0(\xi+\phi_0(t,y))
	+\sum_{i>0}\phi_i(t,y){\bf u}_i(\xi) \;,
\end{equation}
where $ \xi = x-ct $. As is customary, we use an abstract notation  
$\langle \xi |{\bf u}_i \rangle = {\bf u}_i(\xi)$. The vectors 
$|{\bf u}_i \rangle$ are the solutions of the eigenvalue problem,
\begin{equation}
\label{eigen}
\displaystyle \left[{\cal D}\hat{{\bf F}}{\bf (z_0)} + {\bf D}
	\frac{\partial^2}{\partial \hat{\xi}^2} +
	c \frac{\partial}{\partial \hat{\xi}}\right]|{\bf u}_i\rangle
	\equiv \hat{{\cal L}}|{\bf u}_i\rangle
	= \lambda_i|{\bf u}_i\rangle \;,
\end{equation}
where  ${\cal D} {\bf F}$ is the Jacobian of the vector-function $\bf F$ and  
$\hat{\cal L}$ is the operator in square brackets. Here a hat
signifies an abstract operator while the corresponding quantity 
without a hat is its $\xi$ representation. 
We expand $ {\bf F(\bf z)} $ in a Taylor series near $ {\bf z}_0 $
and neglect terms of second order and higher. Since the main features 
of the front dynamics are embodied in ${\cal D} {\bf F}$, 
the truncation of the Taylor series does not discard any important information.   
Due to the translational symmetry of equation (\ref{reaction})
we have a straightforward solution of (\ref{eigen}), namely
\mbox{$\displaystyle |{\bf u}_0\rangle = \partial 
|{\bf z}_0 \rangle/\partial \hat{\xi}$}
corresponding to a zero eigenvalue. Taking this solution
into account we write
\begin{equation}
\begin{array}{rcl}
	\displaystyle {\partial \phi_i \over \partial t}|{\bf u}_i\rangle  
	& = & \displaystyle c|{\bf u}_0\rangle +
	\hat{{\bf F}}({\bf z}_0) +
	{\bf D}\frac{\partial^2}{\partial \hat{\xi}^2}{\bf z}_0  +
	\left(\nabla \phi_0\right)^2
	{\bf D}\frac{\partial}{\partial \hat{\xi}}|{\bf u}_0\rangle + \\
	& & \displaystyle + \Delta \phi_i {\bf D} |{\bf u}_i\rangle +
\sum\limits_{i>0}\phi_i \left[
	{\cal D}{\bf F}|{\bf u}_i\rangle + {\bf D}
	\frac{\partial^2}{\partial \hat{\xi}^2}|{\bf u}_i\rangle +
	c \frac{\partial}{\partial \hat{\xi}}|{\bf u}_i\rangle \right]\;.
\end{array}
	\label{main}
\end{equation}
Using the identity $\displaystyle -c |{\bf u}_0\rangle = {\bf F}({\bf z}_0)
+
{\bf D}\frac{\partial^2}{\partial \hat{\xi}^2}|{\bf z}_0\rangle$ and definition
(\ref{eigen}) we arrive at the expression
\begin{equation}\label{final}
	{\partial \phi_i \over \partial t} |{\bf u}_i\rangle = 
	\lambda_i \phi_i |{\bf u}_i\rangle +
	\Delta \phi_i {\bf D} |{\bf u}_i\rangle + \left(\nabla \phi_0\right)^2
	{\bf D}\frac{\partial}{\partial \hat{\xi}}|{\bf u}_0\rangle \;.
\end{equation}
Here and below we use the Einstein summation convention. 
Alternatively, after multiplication by $\langle{\bf u}_i|$,
(\ref{final}) may be represented by a set of coupled
equations
\begin{equation}\label{eqset}
{\partial \phi_i \over \partial t}  =  
\lambda_i \phi_i + \langle{\bf u}_i|{\bf D}|{\bf u}_j\rangle \Delta
\phi_j +
	\langle{\bf u}_i|{\bf D}\frac{\partial}{\partial \hat{\xi}}|{\bf u}_0\rangle
	\left(\nabla \phi_0\right)^2\;.
\end{equation}

This set of equations can be formally solved 
for modes $ \phi_i $ ($ i \ne 0 $) by treating $ \phi_0 $ as an independent 
function and applying Duhamel's principle to the resulting system of 
inhomogeneous linear equations. We find
\begin{equation}\label{Duham}
	\mbox{\boldmath$\phi$}(t) = e^{\widehat {\bf W} t} \mbox{\boldmath$\phi$}(0) + 
	\int\limits^{t}_{0} ds
	e^{\widehat {\bf W} (t-s) }  \left[
	{\bf a} \Delta \phi_0(s) + {\bf b} \left(\nabla \phi_0(s)\right)^2 \right]\;.
\end{equation}
In (\ref{Duham}) we use the following notation: the matrix operator 
${\widehat {\bf W}}$ has elements 
$ {\widehat W}_{ij} = \delta_{ij} \lambda_i + \langle{\bf u}_i|{\bf D}|{\bf
u}_j\rangle\Delta $, the vectors ${\bf a}$ and ${\bf b}$ have elements 
$ a_i = \langle{\bf u}_i|{\bf D}|{\bf u}_0\rangle $,
$ b_i = \langle{\bf u}_i|{\bf D}\displaystyle\frac{\partial}{\partial \hat\xi}
|{\bf u}_0\rangle $, respectively, and $\mbox{\boldmath$\phi$}$ is a vector with 
elements $\phi_i$ $(i > 0)$. 

Assuming that $ \exp(\widehat {\bf W} t) $ decays rapidly compared to $\phi_0$ 
we may perform the integration in (\ref{Duham}) and substitute the
result into the equation for $ \phi_0 $ to obtain
\begin{equation}\label{complete}
{\partial \phi_0 \over \partial t} = 
\left(a_0 - \sum_{i,j>0} c_i \{ \widehat {\bf W}^{-1}\}_{ij} a_j \Delta \right)
\Delta \phi_0 + \left( b_0 - \sum_{i,j>0}
c_i \{ \widehat {\bf W}^{-1}\}_{ij} b_j \Delta \right)\left(\nabla \phi_0\right)^2\;,
\end{equation}
where $ c_i = \langle{\bf u}_0|{\bf D}|{\bf u}_i\rangle $. This is the 
generalized amplitude equation 
which will form the basis of the analysis of the biscale front chaos 
given below. 

If the dynamics is described by small $ k $ modes one can take $\widehat {\bf W}^{-1}$ 
to be a diagonal matrix $ \delta_{ij}/\lambda_i $ and by omitting second order
terms in
the coefficient of $ (\nabla\phi)^2 $ one obtains\cite{foot1} 
\begin{equation} \label{adiab}
{\partial \phi_0 \over \partial t} = 
\langle{\bf u}_0|{\bf D}|{\bf u}_0\rangle \Delta \phi_0 +
\langle{\bf u}_0|{\bf D}\frac{\partial}{\partial \hat{\xi}}|{\bf u}_0\rangle
\left(\nabla \phi_0\right)^2 - \sum_{i>0}
\frac{\langle{\bf u}_0|{\bf D}|{\bf u}_i\rangle\langle{\bf u}_i|{\bf D}
|{\bf u}_0\rangle}{\lambda_i} \Delta^2 \phi_0 \;.
\end{equation}
The coefficient of the  $\left(\nabla \phi_0\right)^2$
term is obtained from the following identity:
\begin{eqnarray}
0 & \equiv & \int\limits_{-\infty}^{x}\left[
\langle{\bf u}_0|\hat{{\cal L}}{\bf | \xi \rangle}{\bf \langle \xi}|{\bf
u}_0\rangle -
\langle{\bf u}_0|{\bf \xi \rangle}{\bf \langle \xi |}\hat{{\cal L}}|{\bf
u}_0\rangle
\right]d\xi  \nonumber \\
& = & \left.\left[2\langle{\bf u}_0|{\bf \xi\rangle}{\bf \langle\xi|}
{\bf D}\frac{\partial}{\partial \hat{\xi}}|{\bf u}_0\rangle +
\left(c - {\bf D}\frac{\partial}{\partial \xi}
\right)\langle{\bf u}_0|{\bf \xi\rangle\langle\xi}|{\bf u}_0\rangle
\right]\right|_x \;.
\end{eqnarray}
The above formula vanishes for all $ x $ so that after integrating
over the entire domain we obtain
\mbox{$\displaystyle \langle{\bf u}_0|{\bf D}\frac{\partial}{\partial \hat{\xi}}|{\bf
u}_0\rangle
= -\frac{c}{2}$} and observe that the expression has a universal
character. This gives the Kuramoto-Sivashinsky equation,
\begin{equation} \label{kseq}
{\partial \phi_0 \over \partial t} = 
\nu \Delta \phi_0 
-{c \over 2} \left(\nabla \phi_0\right)^2 - 
\kappa \Delta^2 \phi_0 \;,
\end{equation}
with 
\begin{equation}
\nu=\langle{\bf u}_0|{\bf D}|{\bf u}_0\rangle\;, 
\label{nueq}
\end{equation}
and
\begin{equation}
\kappa=\sum_{i>0} \frac{\langle{\bf u}_0|{\bf D}|{\bf u}_i\rangle\langle
{\bf u}_i|{\bf D} |{\bf u}_0\rangle}{\lambda_i}\;.
\label{kappaeq}
\end{equation}
A stability analysis of (\ref{eqset})
shows that the planar front is unstable with respect to long-scale, 
small-amplitude perturbations when $\nu<0$.

In the next subsection we analyse the cubic autocatalysis front dynamics in 
terms of both the generalized amplitude equation (\ref{complete}) as well 
as the Kuramoto-Sivashinsky equation (\ref{kseq}).

\subsection{Amplitude equation description of front dynamics}

\subsubsection{Dynamics near instability} 
The Kuramoto-Sivashinsky equation provides
insight into the nature of the cubic autocatalysis fronts close to the
instability point, and we briefly comment on this connection. 
Simulations of the Kuramoto-Sivashinsky equation produce height correlation
functions and space-time plots (cf. Fig.~\ref{fig:spagen}) of the front 
dynamics similar to those of the cubic autocatalysis model.  
In terms of the Kuramoto-Sivahinsky equation, one sees that the front dynamics for 
$\delta > \delta_c$ is determined by the instability at long wavelengths,
due to the fact that $\nu < 0$ in the $\Delta \phi_0$ term, 
and the dissipation at short wavelengths controlled by the $\Delta^2 \phi_0$ 
term with positive $\kappa$. The dispersion relation of the Kuramoto-Sivashinsky 
equation is $\omega(k)=-\nu k^2 -\kappa k^4$ with maximum at $k_m=(-\nu/2
\kappa)^{1/2}$. The long wavelength structure drives the
dynamics of the extrema (``particles") whose collision dynamics provides
the dissipation of the energy. The characteristic distance between
extrema can be related to the wavevector by $\ell^*=2 \pi/k_m$. Using 
a zeroth order approximation to the $\nu$ and $\kappa$ coefficients given 
in the next section, for $\delta=5$ we obtain $\ell^*=30$ which is
is comparable to the value of $\ell^*=39$ found in the numerical simulations.
Using the methods given in Sec.~\ref{sec:params} it is possible to obtain
$\nu$ and $\kappa$ accurately and improve this estimate. For the present
purposes this rough estimate plus the appearance of the space-time 
plot\cite{pik:bohr}
in Fig.~\ref{fig:spagen} suffice to confirm the general character
of the Kuramoto-Sivashinsky mechanism for the chaotic cubic autocatalysis 
front dyanmics for large system lengths.

It has been argued\cite{yak:lar} that the long-scale dynamics of the 
one-dimesional Kuramoto-Sivashinsky equation can be described by the 
Kardar-Parisi-Zhang equation\cite{kpz},
\begin{equation}
{\partial \phi_0(t,y) \over \partial t}= {\cal D} 
{\partial^2 \phi_0(t,y) \over \partial y^2} +{c \over 2} (\nabla \phi_0(t,y))^2
+\xi(t,y)\;,
\label{kpzeq}
\end{equation}
where $c$ is the front speed, ${\cal D}$ is a diffusion coefficient and 
$\xi(t)$ is a Gaussian white noise source with correlation function 
$\langle \xi(t,y)\xi(t',y') \rangle =2\Gamma \delta(t-t')\delta(y-y')$.
Although there is debate about some aspects of the reduction in higher 
dimensions\cite{pro:red}, numerical simulations\cite{sne:dyn} of the scaling 
properties of the Kuramoto-Sivashinsky equation have confirmed this reduction in 
one dimension. The diffusive character of the cubic autocatalysis front
dynamics discussed in Sec.~\ref{sec:phen} is consistent with such a 
description and provides additional confirmation of 
the applicability of the Kuramoto-Sivashinsky equation close to the onset of
instability.

\subsubsection{Biscale chaos far from instability}
Far from instability, for $\delta \gg \delta_c$, we have seen that the
cubic autocatalysis model presents new features. The
space-time plot (Figure~\ref{fig:min}(b)) shows that besides the relatively
short-lived trajectories, which are similar to those of the
Kuramoto-Sivahinsky equation, there exist long-lived, stable patterns. We have
verified that
this type of biscale chaotic front dynamics, which involves a second
characteristic length scale, cannot be
described by the Kuramoto-Sivashinsky equation. One may seek a generalization
of
the Kuramoto-Sivashinsky
equation in which the coefficients of the $\Delta \phi_0$ and
$\Delta^2 \phi_0$ terms are modified; this in effect changes the
dispersion relation and if it has the form sketched in Fig.~\ref{fig:disp}(a)
with two extrema then one might observe biscale spatio-temporal dynamics.
We note, however, that direct numerical
calculation of the fastest growth
mode and the growth rate corresponding to it (cf. Appendix for the
method used to carry out this calculation) shows that the
dispersion relation remains similar to that of the Kuramoto-Sivashinsky
equation even for
$\delta \gg \delta_c$ (see Fig.~\ref{fig:disp} for a schematic representation
of the situation).
Thus one must seek a description of this new type of front dynamics outside the
usual Kuramoto-Sivashinsky equation or modifications of it that involve a
generalized dispersion relation.

We now describe a model that
captures the qualitative features of this phenomenon and is consistent with the
amplitude equation (\ref{complete}) derived above. Since
the origin of the new dynamics does not reside in terms that affect
the dispersion relation, one is led to examine the nonlinear term in
(\ref{complete}) which, for future reference, we 
rewrite in terms of its Fourier transform:
\begin{equation}\label{agenks}
	\frac{\partial \phi_0(t,k)}{\partial t} =
	\omega(k)\phi_0(t,k) + \frac{c g(k)}{4\pi}\sum
	\limits_{k'} k'(k-k')\phi_0(t,k')\phi_0(t,k-k') \;,
\end{equation}
where $ g(k) $ is given by \begin{equation}
    -{c \over 2} g(k) = b_0 -
	\sum_{i,j>0}c_i \{ \widehat {\bf W}(k)^{-1}\}_{ij} b_j k^2\;.
\end{equation}
Recall that $b_0=-c/2$. The dynamical system governed by
$ \partial \phi/\partial t =(\nabla\phi)^2$ conserves the
quantity $\displaystyle \int(\nabla\phi)^2 d\xi $ as can be seen  easily
from the identity\cite{zale:stoc}:
\begin{equation}
	\frac{d}{d\tau} \int(\nabla\phi)^2 d\xi =  - 2\int \nabla^2\phi
	(\nabla\phi)^2 d\xi = -\frac{2}{3} \int\nabla(\nabla\phi)^3
	d\xi \;,
\end{equation}
so that below we shall refer to the linear and nonlinear terms in
(\ref{agenks}) as dissipation and mixing terms, respectively.
Multiplying (\ref{agenks}) by $ \phi_0(t,-k) $ and averaging the
resulting equation over time we have for the stationary power spectrum
\begin{equation}\label{pwr}
	0=\omega(k) E(k) + \frac{c g(k)}{4\pi}\sum
	\limits_{k'} k'(k-k') \langle \phi_0(t,k')\phi_0(t,k-k')
	\phi_0(t,-k) \rangle \;.
\end{equation}
From this equation it is clear that the stationary properties depend on
the ratio $\omega(k)/g(k)$ so that for $ g(k) $ as in
Fig.~\ref{fig:disp}(b) we see that the growth of the mode with wavenumber
$ k_{0} $  is strongly suppressed. While the above arguments do not constitute
a rigorous ``proof'' they serve as a guide to the 
proper form of the mixing coefficient. Technically,
the above scheme is accomplished in a natural way
when a zero of $ \widehat {\bf W}(k) $ comes
close to real axis in the vicinity  of the maximum of the
dispersion relation. In this case behaviour of $g(k)$ is dominated
by the pole of $ \widehat {\bf W}^{-1}(k) $ and the mixing coefficient
may be approximated by
\begin{equation}
	g(k) = 1 +   \frac{\epsilon k^2}{(k-k_0)^2+\gamma^2}  \;,
\end{equation}
where the damping term can be rationalized in terms of the
continuous spectrum of $ \lambda_i $.

In the simulation of this equation we scale space, time and $\phi_0$ so
that $\nu$, $\kappa$ and $c/2$ are unity and take the following
values for the parameters that appear in $g(k)$:
$ \epsilon=0.2 $, $ k_0=0.6 $ and $ \gamma^2=0.005 $. The results are
presented in Fig.~\ref{fig:spagen}(b).
In the figure one sees the existence of the structures with
two different length scales, the main signature of biscale chaos. The 
generalized amplitude equation does not capture all of the details of 
cubic autocatalysis front dynamics far beyond the onset of instability. 
For example, the simulations of cubic autocatalysis show a broader distribution 
of longer length scales and, as is quite evident in Fig.~\ref{fig:min}, 
regions where there are no extrema. The full descriptions of these 
features may be specific to cubic autocatalysis and require a more detailed 
analysis. However, the main feature, the emergence of a second length 
scale, is described by the model.

We note the similarities in transitions to
ordinary and biscale chaotic regimes. In the  former case the transition
takes place when \mbox{$\langle{\bf u}_0|{\bf D}|{\bf
u}_0\rangle$} passes through zero and in the latter
we require a zero of \mbox{$\left\{\delta_{ij} \lambda_i - \langle{\bf
u}_i|{\bf D}|{\bf
u}_j\rangle k^2\right\}$} to cross the real axis.
The correspondence between the pole of $ \widehat {\bf W}^{-1} $ and the
existence of structures with different wavenumbers suggests that
this phenomenon can arise in other contexts.
The existence of more complicated structures, say
with three or four different wavenumbers, is
unlikely in view of the continuity
of the spectrum and the fact that the dispersion relation
is a smooth function of $k$. Consequently, the amplitude equation
(\ref{agenks}) could serve as the basis for the analysis of such phenomena
in systems other than the cubic autocatalysis reaction.

\section{Parameters in Kuramoto-Sivashinsky Equation} \label{sec:params}
Thus far we have not related the parameters in the Kuramoto-Sivashinsky equation 
to those in the cubic autocatalysis model (\ref{rdeq}) and in this
section we determine $\nu$ and $\kappa$. 
We first present an approximate analytical calculation of these 
parameters based on the exact solutions of the right and left eigenvalue 
problems for $|{\bf u}_0 \rangle $and $\langle {\bf u}_0|$, respectively, 
for case of equal diffusion coefficients\cite{koga:loca}. This provides
insight into the structure of the eigenvalue spectrum and the
corresponding eigenvalues. 
We then describe a general numerical scheme which  can be applied
to other models as well. In this section we choose a
different space scaling and set $D_A=1+\mu$ and $D_B=1-\mu$ since this 
simplifies the computations. 

\subsection{Case of equal diffusion coefficients}
When the diffusion coefficients are equal ($ \mu = 0 $) 
the reaction-diffusion equation (\ref{rdeq}) supports an additional 
relation involving the $\alpha$ and $\beta$ concentrations due to conservation
of the total mass at each point of space, $\alpha+\beta=1$, given that at 
the initial time this relation is satisfied at every space point. 
We can rewrite system (\ref{rdeq}) ($D_A=D_B=1$) in the form
\begin{equation}
{\partial \alpha \over \partial t} = -\alpha(1-\alpha)^2 + \Delta\alpha \;. 
\end{equation}
Local initial input of the autocatalyst induces the creation of a front,
propagating with the minimal velocity \cite{bill:deve}, whose profile is
\begin{equation}  %\label{alpprof}
	\alpha_0(\xi) = (1+\exp(-c\xi))^{-1} \mbox{ , }
\end{equation}
where $\displaystyle c = 1/\sqrt{2}$ is the velocity of the front. The
form of the $\beta$ profile follows from the mass conservation relation.
Henceforth, when confusion is unlikely to arise, we drop the subscript
$0$ on $\alpha$ and $\beta$ to simplify the notation.

Next we discuss the spectrum of the eigenvalue problem (\ref{eigen})
for the cubic autocatalysis problem when the diffusion coefficients are
equal. Using the explicit form of ${\cal D}{\bf F}({\bf z}_0)$ 
derived from (\ref{rdeq}) we have 
\begin{eqnarray}
[{\partial^2 \over \partial \xi^2} +c {\partial \over \partial \xi}-\beta^2] 
X_1 -2\alpha \beta X_2&=&\lambda X_1 \;,\nonumber \\
 \beta^2 X_1 
 +[{\partial^2 \over \partial \xi^2} +c {\partial \over \partial \xi}+2\alpha \beta]
X_2 &=&\lambda X_2\;,
\label{eqeval}
\end{eqnarray}
where we have written the components of a general right eigenvector ${\bf u}$ as 
${\bf u}=(X_1,X_2)$. We shall show
that the spectrum is obtained from two distinct equations.
First, by adding the two lines
of (\ref{eqeval}), we obtain a linear differential equation with
constant coefficients: 
\begin{equation}
 \left[
\frac{\partial^2}{\partial \xi^2} + c \frac{\partial}{\partial \xi}
\right]X = \lambda X \;,
\end{equation}
where $X=X_1+X_2$. After a change of variables
$\displaystyle X =  \exp(-c\xi/2)Y $ we obtain
$ \displaystyle \left[\partial^2/\partial \xi^2-c^2/4\right]Y = \lambda Y $.
The eigenvalue problem for $Y$ has a continuous spectrum
$\displaystyle \left(-\infty,-c^2/4\right]$.
In these considerations we have omitted the case $ X_1+X_2 = 0 $, which
satisfies the eigenproblem for $X$ automatically.
Using this relation we obtain the following equation for the other part of 
the spectrum:
\begin{equation}\label{alph}
\left[ \frac{\partial^2}{\partial \xi^2} + c \frac{\partial}{\partial \xi}
+ (2\alpha\beta - \beta^2)\right] X_1 = \lambda X_1\;.
\end{equation}
After the same change of variables 
$\displaystyle X_1 =  \exp(-c \xi/2) Y_1$ we get 
a ``Schr\"odinger" equation with
potential \mbox{$ V = -2\alpha\beta + \beta^2$} (cf.
Fig.~\ref{fig:wkbpot}(a))  
and energy levels \mbox{$\displaystyle E_i = -c^2/4 - \lambda_i $}.
We see that this equation has a continuous spectrum
starting from zero as well as discrete spectrum. The WKB method may be used 
to obtain approximations for the eigenvalues and eigenfunctions. Applying the
Bohr-Sommerfeld quantization rule we get $E_0 = -0.1066$ for the first 
energy level, which is in good agreement with the theoretical
value $ -c^2/4=-0.125 $. This indicates that the  WKB approximation is 
reliable and it predicts that the potential well is too
shallow\cite{foot2} to have other discrete energy levels.
So we conclude that the spectrum consists of a zero eigenvalue
and doubly degenerate continuous spectrum lying in
$\displaystyle \left(-\infty,-c^2/4 \right]$.

The right eigenvector \mbox{$\displaystyle |{\bf u}_0\rangle = 
\partial |{\bf z}_0 \rangle/\partial \hat{\xi}$} was 
discussed in Sec.~\ref{sec:phen}, but the left eigenvector $\langle{\bf u}_0|$ is more 
difficult to obtain. We now construct the explicit form for $\langle{\bf u}_0|$ 
for the case of equal diffusion. Letting 
$\langle{\bf u}_0|\xi \rangle={\bf u}_0^*(\xi)=(X_1^*,X_2^*)$ the eigenvalue 
problem for ${\bf u}_0^*(\xi)$ takes the form
\begin{equation}
\label{fleig}
\begin{array}{rcl}
 \displaystyle \left[ \frac{\partial^2}{\partial \xi^2} -
c \frac{\partial}{\partial \xi} \right] X_1^* - \beta^2 (X_1^*-X_2^*) & = & 0\;, 
\nonumber \\
 \displaystyle \left[\frac{\partial^2}{\partial \xi^2} -
c \frac{\partial}{\partial \xi} \right] X_2^* - 2\alpha\beta
(X_1^*-X_2^*) & = & 0 \;,
\end{array}
\end{equation}
so that the difference $ X^* = X_1^* - X_2^* $ obeys
\begin{displaymath}
\left[ \frac{\partial^2}{\partial \xi^2} -
c \frac{\partial}{\partial \xi} + (2\alpha\beta - \beta^2)\right] X^* = 0\;.
\end{displaymath}
This can be cast into the form of (\ref{alph}) 
after the change of variable \mbox{$ X^* = A \exp(c\xi) Z $},
where $ A $ is some constant defined by the 
normalization  condition $ {\bf \langle u}_0{\bf | u}_0{\bf \rangle}=1 $:
\begin{equation}
	{\bf \langle u}_0{\bf | u}_0{\bf \rangle} = \int
\limits_{-\infty}^{\infty}
	\frac{\partial \alpha}{\partial \xi} X^*
	d \xi = c A \int \limits_{0}^{1} \alpha^2 d \alpha = 
	\frac{c A}{3}\;;
\end{equation}
hence, we have 
\begin{equation}
	\label{fdif}
	X^* = 3(1+\exp(-c\xi))^{-2} \;.
\end{equation}
Performing the change of variable $\Psi = \exp(-c\xi/2) X_1^*$ in
(\ref{fleig}) and using (\ref{fdif}) we obtain 
\begin{equation}
	\label{fphi1}
	\frac{\partial^2}{\partial \xi^2} \Psi - \left(\frac{c}{2}\right)^2
\Psi
	= 3\beta^2\exp(-c\xi/2)(1+\exp(-c\xi))^{-2}\;.
\end{equation}
The Green's function for the operator $\displaystyle \left[
\partial^2/\partial \xi^2 - c^2/4 \right] $ is given
by $ \displaystyle {\cal G}(\xi) = -c^{-1}\exp(-c|\xi|/2) $.
Convolution of $ {\cal G} $ with the right-hand-side of (\ref{fphi1})
yields a solution for $  X_1^* $,
\begin{equation}
 X_1^*(\xi)  = -\frac{1+2\exp(-c\xi)}{(1+\exp(-c\xi))^2}\;,
\end{equation}
and the difference between $ X_1^*$ and $ X^*$ gives $ X_2^*$, 
\begin{equation}
 X_2^*(\xi)  = -\frac{4+2\exp(-c\xi)}{(1+\exp(-c\xi))^2}\;.
\end{equation}
These functions are plotted in Fig.~\ref{fig:wkbpot}(b). The functions 
have limiting behaviour \
\mbox{$ X^*_1(+\infty)=-1 $} and \mbox{$ X^*_2(+\infty)=-4 $}.

System (\ref{fleig}), for any diffusion coefficient ratio, has another 
zero left eigenfunction ${\bf V}^*=(1,1)$, which
appears due to conservation of the total mass of the reagents.
The right eigenfunction ${\bf V}$ corresponding to ${\bf V}^*$ is
\begin{eqnarray}
	 V_1(\xi) & = & \displaystyle \frac{2}{3+3\exp(-c\xi)}\left(
	2+\frac{2c\xi+3}{1+\exp(c\xi)} \right)\;,  \\
	V_2(\xi) & = & 1 - \displaystyle \frac{2}{3+3\exp(-c\xi)}\left(
	1+\frac{2c\xi+3}{1+\exp(c\xi)} \right)  \;.
\end{eqnarray}
While both the ``bra" vector corresponding to spatial
invariance and the ``ket" vector obtained above do not decay to zero at
$\xi=\infty$ their scalar product does and the product is given,
after change of variable, by:
\begin{equation}
	\langle {\bf V} | {\bf u}_0 \rangle =
	\sqrt{2} \int\limits^{1}_{0} \left[ 4\alpha^2 \ln\frac{\alpha}
	{1-\alpha} - 2(1+3\alpha)(1-\alpha)\right]d\alpha = 0\;.
\end{equation}
Since the above integral vanishes these considerations insure us that
we have an orthogonal system of ``bra" and ``ket" vectors.
Moreover, the fact that one product is finite and the other
is not removes an ambiguity in the definition of the left eigenfunction.
We believe the same situation holds for arbitrary $\mu$.

\subsubsection{Estimate of $\nu$ for $\mu \ne 0$} 
For equal diffusion coefficients ${\bf D}$ is the identity matrix and it 
follows that $\nu= \langle{\bf u}_0|{\bf D}|{\bf u}_0\rangle = 1$. 
However, using $ |{\bf u}_0\rangle $ and $\langle{\bf u}_0|$ computed
for $\mu=0$, we may obtain a first order approximation for $\nu$ for
unequal diffusion coefficients. From (\ref{nueq}) we write:
\begin{equation}\label{inst}
	\nu=\langle{\bf u}_0|{\bf D}|{\bf u}_0\rangle =
	1 + \mu \langle{\bf u}_0|{\mbox{\bf I}}|{\bf u}_0\rangle \;,
\end{equation}
where $ \displaystyle
	\mbox{\bf I} = \left( \begin{array}{cc}
	1 & 0 \\ 0 & -1 \end{array} \right)  $.
Using the explicit, equal-diffusion forms for $\langle{\bf u}_0| $ and 
$ |{\bf u}_0\rangle $ we get
 \begin{equation}
 \langle{\bf u}_0|{\mbox{\bf I}}|{\bf u}_0\rangle = \int
\limits_{-\infty}^{\infty}
 \frac{\partial \alpha}{\partial \xi} \left(X^*_1 + X^*_2\right)
	d \xi = - \int \limits_{0}^{1} \left( \alpha^2 +
	4 \alpha \right) d \alpha = -\frac{7}{3} \;,
\end{equation}
and $\nu=1-7\mu/3$. 

\subsubsection{Estimate of $\kappa$ for $\mu \ne 0$}
When the diffusion coefficients are equal and {\bf D} is the identity
matrix we have $\kappa=0$ in view of the orthogonality of the
eigenfunctions. However, as in the preceding subsection we can obtain an
estimate for $\kappa$ for arbitrary $\mu$ using the solution of the
eigenvalue problem for $\mu=0$. To obtain this estimate we 
assert completeness of the set of eigenfunctions; i.e., \mbox{$ \displaystyle
\sum\limits_{i} |{\bf u}_i\rangle\langle{\bf u}_i| = {\bf 1}$}.
Using this assumption we obtain\cite{foot3}
\begin{equation}
	\sum\limits_{i} \frac{\langle{\bf u}_0|A|{\bf u}_i\rangle\langle{\bf
u}_i|B|{\bf u}_0\rangle}{\lambda_i}
	= \langle{\bf u}_0|B|{\bf x}\rangle-\langle{\bf u}_0|B|{\bf
u}_0\rangle\langle{\bf u}_0|{\bf x}\rangle \;,
\label{keq}
\end{equation}
where $ |{\bf x}\rangle $ is a solution of the equation
\begin{equation}\label{trdtrm}
	\left[{\cal D}{\bf F} + {\bf D}
	\frac{\partial^2}{\partial \xi^2} +
	c \frac{\partial}{\partial \xi}\right]|{\bf x}\rangle =
	A|{\bf u}_0\rangle-\langle{\bf u}_0|A|{\bf u}_0\rangle |{\bf
u}_0\rangle \;.
\end{equation}
In this gereral expression we let $A=B={\bf D}$ and use  
$ \langle{\bf u}_0|{\bf D}|{\bf u}_i\rangle
\langle{\bf u}_i|{\bf D}|{\bf u}_0\rangle = \mu^2\langle{\bf u}_0|{\bf I}|{\bf
u}_i\rangle
\langle{\bf u}_i|{\bf I}|{\bf u}_0\rangle $ in (\ref{trdtrm}). Using 
$A={\bf D}$ and adding the two lines of (\ref{trdtrm}) we have
\begin{equation}
	\left[ \frac{\partial^2}{\partial \xi^2} +
	c \frac{\partial}{\partial \xi}\right] (x_1+x_2) = 2\alpha_{\xi} \;,
\end{equation}
with $\alpha_{\xi}=d\alpha/d\xi$. Solving the above equation with appropriate 
boundary conditions we get
\begin{equation}
	x_1 + x_2 = - \frac{2}{c} \frac{\displaystyle \log(1+e^{c\xi})}
{\displaystyle e^{c\xi} } \;.
\end{equation}
Equation (\ref{trdtrm}) can be solved for $ x_1 $ to obtain (we use in the next
equation the explicit form for $ \langle{\bf u}_0|{\bf I}|{\bf u}_0\rangle $)
\begin{equation}\label{solut}
	\left[ \frac{\partial^2}{\partial \xi^2} +
	c \frac{\partial}{\partial \xi} + 2\alpha\beta -
	\beta^2 \right] x_1 =  2\alpha\beta (x_1+x_2) + \frac{10}{3}
	\alpha_{\xi}\;.
\end{equation}
This equation takes an especially simple form after a change of
variable $ \xi \to \alpha $. The derivative possesses the form \mbox{$
\displaystyle \frac{\partial}{\partial \xi} = c\alpha(1-\alpha)
\frac{\partial}{\partial \alpha}$} and (\ref{solut})
after simplifications is tantamount to
\begin{equation}
	\alpha\left (1-\alpha\right ){\frac {\partial^{2}}
	{\partial{\alpha}^{2}}}x_1+2\,\left (1-\alpha
	\right ){\frac {\partial}{\partial\alpha}}x_1+2\,\left
	(3-{\alpha}^{-1}\right )x_1= \frac{8}{c}\left(
	\frac {5}{12}+\frac {\left (1-\alpha\right )
	\ln (1-\alpha)}{\alpha} \right)\;.
\end{equation}
We know that $ \alpha\beta $ is a solution of the homogeneous equation.
By the substitution $ x_1 = \alpha\beta  \chi $ we decrease the order
of (\ref{solut}).
Solving the resulting linear first order equation we obtain the solution
\begin{equation}
	x_1 = \frac{2}{9 c}
	\left(2 \alpha^2 \left (1-{\alpha}^{2}\right )
	\left (\ln (1-\alpha)+\alpha+{\displaystyle
	\frac{\alpha^2}{2}}
	\right )\alpha^2+7\,\alpha\left (1-\alpha\right )
	\ln ({\frac {1-\alpha}{\alpha}})\right)\;.
\end{equation}
From the form of the solution it is clear that the solution decays faster
then $\displaystyle e^{-\gamma c|\xi|} $, $0<\gamma <1$, and thus the scalar 
product is well defined. To integrate we use the identity
\begin{equation}
	\int\limits_{-\infty}^{\infty} Q d \xi =
	\frac{1}{c} \int\limits_{0}^{1} \frac{Q}
	{\alpha(1-\alpha)}d\alpha \;,
\end{equation}
and integration gives 
\begin{equation}
\kappa=\sum\limits_{i > 0} \frac{\langle{\bf u}_0|{\bf D} |
{\bf u}_i\rangle\langle{\bf u}_i|{\bf D}|{\bf u}_0\rangle}
{\lambda_i} = \frac{264 + 40 \pi^2}{27} \mu^2 \;.
\label{kapest}
\end{equation}

Finally, collecting the above information, we arrive at the following 
expression for the Kuramoto-Sivahinsky equation with approximate 
coefficients deduced from the cubic autocatalysis model (\ref{rdeq}): 
\begin{equation}
 {\partial \phi_0 \over \partial t}= (1-\frac{7}{3}\mu)\nabla^2\phi_0 - 
 \frac{264 + 40 \pi^2}{27} \mu^2
	\nabla^4\phi_0 - \frac{c}{2} (\nabla\phi_0)^2\;.
\end{equation}

\subsection{Numerical computation of $\nu$ for $\mu \ne 0$}
In order to determine $\nu$ exactly, 
within the present approach, one must construct a scheme to 
determine the left eigenvector ${\bf u}_0^*$, the solution of 
${\cal L}^{\dagger} {\bf u}^*=\lambda {\bf u}^*$, for $\lambda=0$ and 
arbitrary $\mu$. For the cubic autocatalysis problem the 
left eigenvectors are given by the solutions of 
\begin{eqnarray}
\left[(1+\mu) {\partial^2 \over \partial \xi^2} -c{\partial \over \partial \xi} 
-\beta^2\right]\Psi_1
+\beta^2 \Psi_2 &=& \lambda \Psi_1 \nonumber \\
-2\alpha \beta \Psi_1 +
\left[(1-\mu) {\partial^2 \over \partial \xi^2} -c{\partial \over \partial \xi} 
+2 \alpha \beta\right]\Psi_2
&=& \lambda \Psi_2 \;,
\label{left}
\end{eqnarray}
where we have written a general left eigenvector as 
${\bf u}^*=(\Psi_1,\Psi_2)$.
The solution of this eigenvalue problem in terms of a complete basis 
entails the solution of a difficult 
sparse, non-symmetric, non-tridiagonal matrix eigenvalue problem.  
In order to avoid solving the eigenvalue problem by matrix methods we use
the following device.
We consider an auxiliary problem that originates from the system
of equations (\ref{left}). In (\ref{left}) we replace terms
$\displaystyle c\frac{\partial}{\partial \xi} $ by the time derivative and take
$\alpha$ and $\beta$ to be in a moving frame with
velocity $-c$ (i.e., $ \alpha(t,x) = \alpha_0(x+ct) $ ,
$\beta(t,x) = \beta_0(x+ct)$ ):
\begin{eqnarray}
\label{sysdev}
\displaystyle {\partial \Psi_1 \over \partial t} & = & (1+\mu)
\Delta\Psi_1 - \beta^2(\Psi_1-\Psi_2) \nonumber \\
\displaystyle {\partial \Psi_2 \over \partial t} & = & (1-\mu)
\Delta\Psi_1 - 2\alpha\beta(\Psi_1-\Psi_2) \;.
\end{eqnarray}
Eigenvalues of the adjoint eigenvalue problem
$\displaystyle {\cal L}^{\dagger} {\bf u}^* = \lambda {\bf u}^* $ coincide with those
of (\ref{eigen}) and for nonzero eigenvalues
their real parts are negative and separated from zero.
Using this observation we deduce that all but one of the components 
exponentially
decay so that the persistent solution of the system of partial differential
equations (\ref{sysdev}) is proportional to $\langle{\bf u}_0|$.
It is worth noting here that in order to obtain $\langle{\bf u}_0|$ one should
take unequal initial distributions for $\Psi_1$ and $\Psi_2$ 
in (\ref{sysdev}).

Numerical computations of both (\ref{rdeq}) and (\ref{sysdev})
were carried out using the following hybrid of the 
Runge-Kutta and Crank-Nicholson methods:
\begin{equation}
\label{numer}
\begin{array}{rcl}
\displaystyle \frac{{\bf z}'-{\bf z}(t)}{\tau} & = & \displaystyle 
{\bf D}\Delta
\frac{{\bf z}'+{\bf z}(t)}{2} + {\bf F} \left({\bf z}(t)\right) \;, \\
\displaystyle \frac{{\bf z}(t+\tau)-{\bf z}(t)}{\tau} & = & \displaystyle
{\bf D}\Delta
\frac{{\bf z}(t+\tau)+{\bf z}(t)}{2} + {\bf F}\left(
\frac{{\bf z}'+{\bf z}(t)}{2}\right)\;.
\end{array}
\end{equation}
We use a non-uniform distribution of mesh points
with density $\displaystyle x/(1-x^2)^{-1}$,
where $x$ is in the range $[-1,1]$.
The matrix coefficients and vectors were computed using
a finite element method with functions approximated
by a sum of hat functions. During the simulations
the profiles were monitored and constantly shifted to the
origin. To do this the profiles were interpolated using 
a monotonicity-preserving, piecewise-cubic,
Hermite interpolant to a set of data points\cite{foota}.
For the velocity we use the following relation:
\begin{equation}
c = \int\limits_{-\infty}^{\infty} \alpha \beta^2 d \xi \;.
\end{equation}
Due to the fast decay of the rate of the reaction outside the reaction zone
the infinite range
of the integration can be replaced by the limit mesh points
and the value of the definite integral of the interpolant 
is easily computed. 

The results obtained in simulations for the case
of equal diffusion coefficients were compared with the analytical
results from the previous section. The absolute errors for the
velocity and for $\displaystyle \langle{\bf u}_0|{\bf I}|{\bf u}_0\rangle $
are about $ 10^{-6} $ and $ 10^{-4} $,
respectively.  This confirms that the numerical procedure is
accurate. 

\section{Critical Diffusion Ratio} \label{sec:ratio}
While it is clear that the planar front is unstable for 
sufficiently large values of the diffusion ratio $\delta$, no
estimate was provided the critical diffusion ratio, $\delta_c$, where this
instability takes place. There are a number of estimates of 
$\delta_c$ in the literature. Horv\'ath et al.\cite{horv:inst} estimated
$\delta_c \approx 2.9$ from direct simulation of the reaction-diffusion
equation and found $\delta_c=2$ from an application of Sivashinsky's
method, which assumes an infinitely thin reaction zone, 
to the cubic autocatalysis problem. Zhang and Falle\cite{zhan:stab}
computed the dispersion relation directly and found $\delta_c \approx
2.0$. Finally, Milton and Scott\cite{milt:inst} obtained 
$\delta_c\approx2.366$ using the eikonal theory and $\delta_c \approx 2.25$ 
from numerical simulation. This quantity is difficult to estimate by 
simulation of the reaction-diffusion equation  
and its direct calculation presents a number of interesting features. 

In the context of the amplitude equations discussed in Sec.~\ref{sec:anal},  
$\delta_c$ can be determined from the condition 
$\nu=\langle{\bf u}_0|{\bf D}|{\bf u}_0\rangle=0$. This, in turn, requires
the computation of the left and right eigenvectors of the eigenvalue
problem\cite{foot4}. The analytical estimate of $\nu$ 
discussed above immediately gives an estimate of the critical diffusion 
coefficient ratio. The critical value of $\mu$, $\mu_c$, is given by 
the solution of $\nu=0=1-7\mu_c/3$ or $\mu_c=3/7$. Since 
$\delta=(1+\mu)/(1-\mu)$ we have $\delta_c=2.5$.

To determine the critical diffusion ratio exactly we use the numerical
scheme outlined above to determine the left and right eigenvectors for arbitrary 
$\mu$. In this way the critical diffusion ratio can be obtained iteratively. 
At each step the ratio from the previous run was taken and new profiles and
the ratio $\displaystyle -\langle{\bf u}_0|{\bf I}|{\bf u}_0\rangle^{-1} $
corresponding to them were determined. Iterations rapidly converge to the 
value \mbox{$\mu_c=0.39397$}, which
corresponds to diffusion coefficient ratio $ \delta_c=2.300 $.

We have also computed the dispersion relation to provide additional 
confirmation that $\delta_c >2$.  The details of the computation 
are given in the Appendix. The method we use allows us to put several 
thousand mesh points in the
reaction zone (compared to about 40 in \cite{zhan:stab}). 
In the small wavenumber range the dispersion relation 
should have a dependence on $ k $
given by that of Kuramoto-Sivashinsky equation.  The maximum growth
rate for the dispersion relation \mbox{$ \omega(k)=-\nu k^2 - \kappa k^4
$} is $\displaystyle \nu^2/4\kappa$.  We see that when $ \delta $
approaches $ \delta_{c} $, so that
\mbox{$\nu\sim(\delta_{c}-\delta)$}, the maximal growth rate  rapidly
decreases and can easily fall out of range of the precision of the
calculation.  Moreover,
direct simulation shows the presence of a very slowly decaying tail
of the autocatalyst and its length increases (cf. Appendix) 
when $ k $ approaches zero.  So the calculations  
could also be affected by the limited system size. We also observe 
that least-square 
fittings of the data in Fig.2(a) of \cite{zhan:stab} gives $\nu=.34$ 
and $\kappa=1.7$ indicating stability of the front. 
A further test of our computation of the 
dispersion relation comes from  the first order approximation of $\nu$ 
in the Kuramoto-Sivahinsky equation when $ \mu \approx \mu_c$: $\mu=\mu_c+\delta \mu$. 
In this case we have $ \nu=1+\mu \langle{\bf u}_0|{\mbox{\bf
I}}|{\bf u}_0\rangle = -\displaystyle \delta\mu/\mu_{c} $. 
The results of our computations of the dispersion relation for the case
$ \mu = 0.45 $ ($\delta=2.636$) are given in Fig.~\ref{fig:cdisp}. 
We see that near the
origin the dispersion relation is well approximated by the 
curve $ \omega = 0.142 k^2 $, where $ 0.142 =
\displaystyle \delta\mu/\mu_{c}$. 

\section{Summary and Discussion}
The chaotic dynamics of cubic autocatalysis fronts shows two distinct
kinds of behaviour depending on the diffusion ratio $D_A/D_B$. Close to
the onset of the front instability the transverse structure can be
described in terms of a single length scale. The characteristic length
is associated with the wavelength of the most unstable mode and the
dynamics can be viewed in terms of a particle picture. The
front extrema can be associated with ``particles" that collide and
coalesce and new particles are created to maintain the average particle
number. This dynamical regime was characterized statistically and shown
to be described by the Kuramoto-Sivashinsky equation.

For large values of $D_A/D_B$ a new dynamical regime was found where
the front is characterized by two length scales. We term this type of 
chaotic front dynamics biscale chaos. Biscale chaos cannot be described
by the Kuramoto-Sivashinsky equation and we introduced an amplitude
equation which is a generalization of this equation.
This amplitude equation accounts for a modified nonlinear energy mixing 
among the modes. We identified this kind of behaviour with a pole near the
real axis of a linear operator obtained in the course of the analysis of
the dynamics of small perturbations.  The observed phenomenon may prove to
be quite general in light of the simple prerequisite for it, namely, the
crossing of the real axis by a pole.  However, we believe that the coexistence
of structures with three or more length scales is unlikely since it 
requires the operator to have more complicated properties. Questions about
qualitative effects of this irregular dynamics on microscopic
quantities (e.g., mean front velocity, diffusion coefficient of
the interface, etc.) remain.  Another point to be investigated is the  
possibility of describing the observed
patterns in terms of the dynamics of interacting pulses, which correspond
to small $k$ modes, driven by noise.

An explicit reduction of the  cubic autocatalysis reaction-diffusion equation 
to the Kuramoto-Sivashinsky equation was carried out and the
coefficients that enter in this equation were determined. Both
approximate analytical values as well as a numerical scheme for their
exact computation for arbitrary $D_A/D_B$ were given. The numerical 
method copes with two problems that arise in such calculations, namely, the 
nonlocal form of the zero eigenvector and the degeneracy of the eigenproblem with 
eigenvalue equal to zero, which comes from the conservation laws for the system.  
The algorithm is easy to implement and proved to be reliable. Using these 
results the critical diffusion ratio where the front instability occurs 
was computed and found to be $D_A/D_B=2.300$.

\acknowledgements This work was supported in part by a grant from the 
Natural Sciences and Engineering Research Council of Canada. A. Careta
benefitted from a fellowship of the Programa de Becas de Formaci\'on de 
Personal Investigador en el Extranjero, Ministerio de Educaci\'on y
Ciencia, Spain.
\bigskip

\noindent
\appendix
\section*{Numerical computation of the dispersion relation}
The dispersion relation for the cubic autocatalysis model was 
computed in the following way: The perturbation of the front solution 
of (\ref{reaction}) may be written in the form
\begin{equation}
{\bf z}(t,x,y) = {\bf z}_0(\xi) + \mbox{\boldmath $\chi$}(t,x) 
\exp(\imath k y)\;.
\end{equation}
We assume, when $ k $ is small, that the largest eigenvalue is
separated from the others. Thus, after integration of the equation
\begin{equation} \label{disp}
{\partial \mbox{\boldmath $\chi$} \over \partial t} = 
{\cal D}{\bf F \mbox{\boldmath $\chi$}} + {\bf D}\left({\partial^2 
\over \partial x^2} -k^2 \right)\mbox{\boldmath $\chi$}\;,
\label{perteq}
\end{equation}
for a sufficiently long time, only the most unstable mode survives.
The system (\ref{disp}) is the same as system (4.7) of \cite{zhan:stab}
after proper scaling
and change of variables \mbox{$\chi_{1} = a_1 - \displaystyle \frac{d
a_0}{d x}$},
\mbox{$\chi_{2} = b_1 - \displaystyle \frac{d b_0}{d x}$} in the notation 
of this paper\cite{foot6}. 
After transients the solution of (\ref{disp}) should propagate with the
same velocity as the front and grow with a rate $\omega(k) $ and thus 
we can replace
$ \displaystyle{\partial \mbox{\boldmath $\chi$} \over \partial t} $ with 
$\omega(k)\mbox{\boldmath $\chi$}-c \mbox{\boldmath $\chi$}_{\xi} $.
In \cite{zhan:stab} the eigenvalue problem with $\mbox{\boldmath $\chi$}$ 
as a vector and $\omega(k) $ as eigenvalue was solved. The two approaches
lead to the same result but the latter necessitates solving an eigenvalue
problem for a non-symmetric, sparse matrix that is in general 
a difficult task. The direct integration if the system (\ref{disp}) is more 
accurate and easier to implement. The results of the integration for
$ \mu=0.45 $ are shown in Fig.~\ref{fig:cdisp}.

To decrease the integration time we used as initial conditions for
$\mbox{\boldmath $\chi$}$ the eigenvector obtained from the 
run for $ k - \delta k $
and by employing numerical differentiation we computed $ |{\bf u}_0\rangle $,
which is solution of the eigenvalue problem for $k=0$.

We also note that the autocatalyst exhibits a slowly decaying 
tail whose length increases as $k$ approaches zero.
The length of the tail can be estimated from the equation, 
$\chi_2'' = k^2 \chi_2 - c(1-\mu)^{-1}\chi_2'$, 
which is obtained from the equation governing the dynamics of the
autocatalyst (second member of (\ref{perteq})) by setting the reaction 
terms to zero.  When $ k $ is small the solution exponentially decays as
 $\displaystyle \exp\left(-(1-\mu)k^2 x/c\right) $ and the length
of the tail is of order \mbox{$L_{ta}\sim\displaystyle c[(1-\mu)k^2]^{-1}$}.

\newpage
\begin{figure}
\caption{ Panels (a) and (b) are space-time 
plots of the minima of the autocatalyst ($B$) profiles for the diffusion 
coefficient ratios $ \delta = 5 $ and $ \delta = 8 $, respectively. The
isoconcentration level for these plots is $\beta_r=0.5$. 
The ordinates range from $0$ to $L$, where the system size
is $L=2048$. In the abscissas the initial time is $t=0$ and the final 
time is $t=100\;000$ for (a) and $t=200\;000$ for (b).}
\label{fig:min}
\end{figure}

\begin{figure}
\caption{ The figure shows $C(t,y)$ for $\delta = 5 $ for 
several values of $t$ indicated on the plot, as well as $\bar{C}(y)$ and the 
theoretical value of $\bar{C}(y)=\Gamma y(L-y)/{\cal D}L$ for a 
diffusive interface where
nonlinear terms in the Kardar-Parisi-Zhang equation can be neglected. 
The parameter $\bar{{\cal D}}=\Gamma/{\cal D}$, determined from a fit to the 
$\bar{C}(y)$ data, has the value $\bar{{\cal D}} \approx 0.2$. 
The parameter $\Gamma$ may be estimated independently from the linear 
growth of the $k=0$ Fourier mode. Its value was found to be 
$\Gamma\approx 0.018$. 
Fronts corresponding to the isoconcentration lines $\beta_r=0.5$ were 
used to construct these correlation functions. The system length was
$L=256$ and $C(t,y)$ was determined from an averages over $25$
realizations of the front evolution. In the computation of $\bar{C}(y)$, we used
$t_0 = 25\;000$ and $t_0 + T = 50\;000$.}
\label{fig:corplot}
\end{figure}

\begin{figure} 
\caption{ Power spectra for $\delta=5$ and $\delta=8$. The isoconcentration 
lines for $\beta_r=0.5$ were used to construct these power spectra. The system
length was $L = 2048$ and $E(k)$ was determined from a single run of the front
evolution. For $\delta = 5$ we used $t_0 = 25\;000$ and $t_0 + T = 100\;000$,
while for $\delta = 8$ we used $t_0 = 50\;000$ and $t_0 + T = 200\;000$.}
\label{fig:pspec}
\end{figure}

\begin{figure}
\caption{Space-time plot of the minima of the profiles for (a) the 
Kuramoto-Sivashinsky equation and (b) the
amplitude equation (\protect \ref{agenks}).}
\label{fig:spagen}
\end{figure}

\begin{figure}
\caption{ (a) The solid line is a schematic 
representation of the dispersion
relation for $\delta > \delta_c$. The dotted line shows the 
hypothetical form of the dispersion
relation with two energy sources. (b) Proposed rate of
energy mixing $g(k)$ that produces biscale patterns.}
\label{fig:disp}
\end{figure}

\begin{figure}
\caption{ (a) Plot of the potential in the 
WKB approximation. (b) Plots of the components of the left eigenvector for
the case of equal diffusion coefficients. }
\label{fig:wkbpot}
\end{figure}

\begin{figure}
\caption{Dispersion relation of the cubic autocatalysis model for the case
\mbox{$ \mu = 0.45 $} (or $\delta= 2.636$). 
The symbols `+' denote the dispersion relation 
constructed from the numerically obtained 
growth rate. It is compared with the first order approximation
\mbox{ $ \omega(k)\approx 0.142 k^2 $} for small values of $ k $.}
\label{fig:cdisp}
\end{figure}

%\newpage
%\begin{figure}[h]
%\epsfxsize = \textwidth
%\epsfbox{fig1.eps}
%\end{figure}
\newpage
\begin{figure}[h]
\epsfxsize = \textwidth
\epsfbox{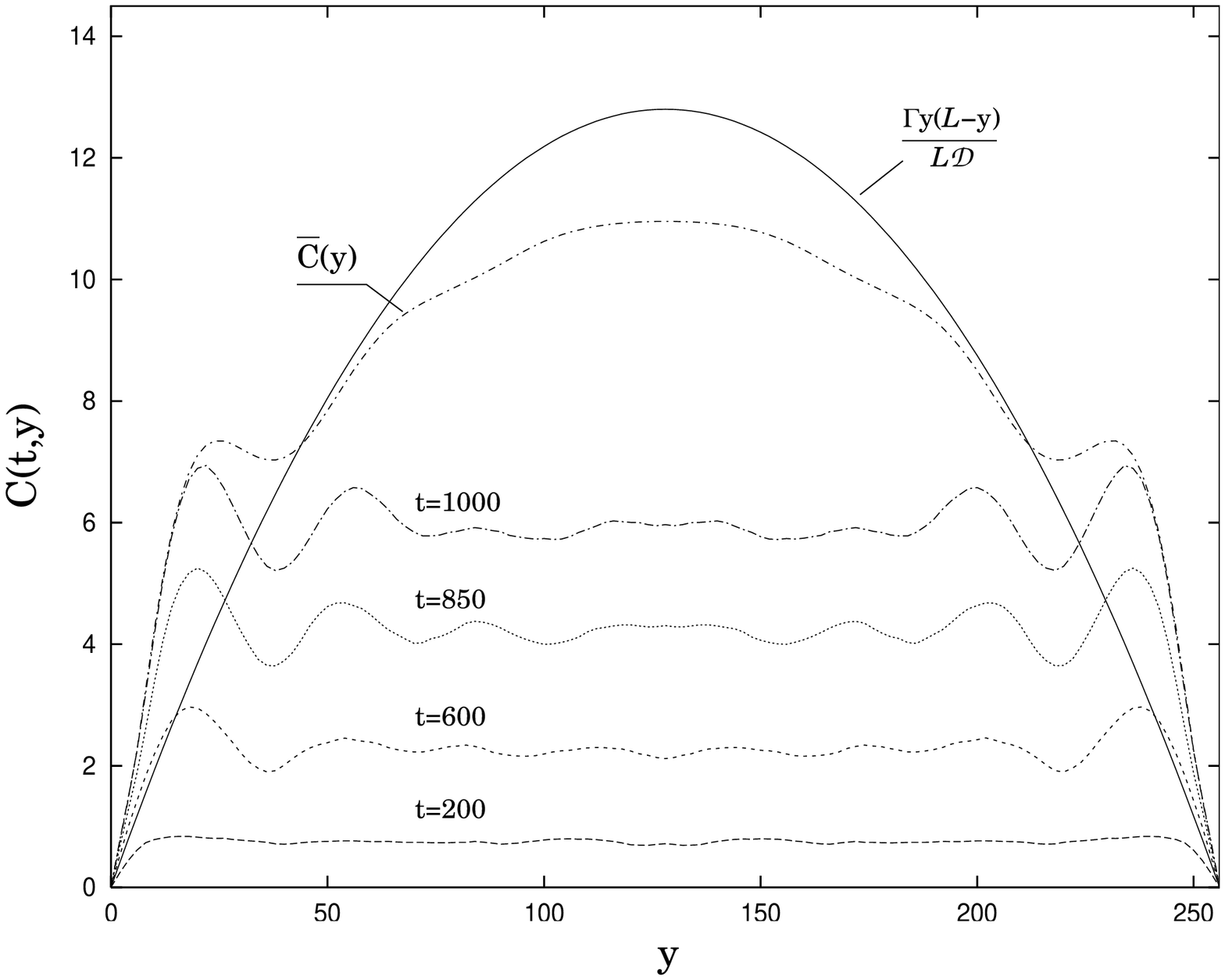}
\end{figure}
\newpage
\begin{figure}[h]
\epsfxsize = \textwidth
\epsfbox{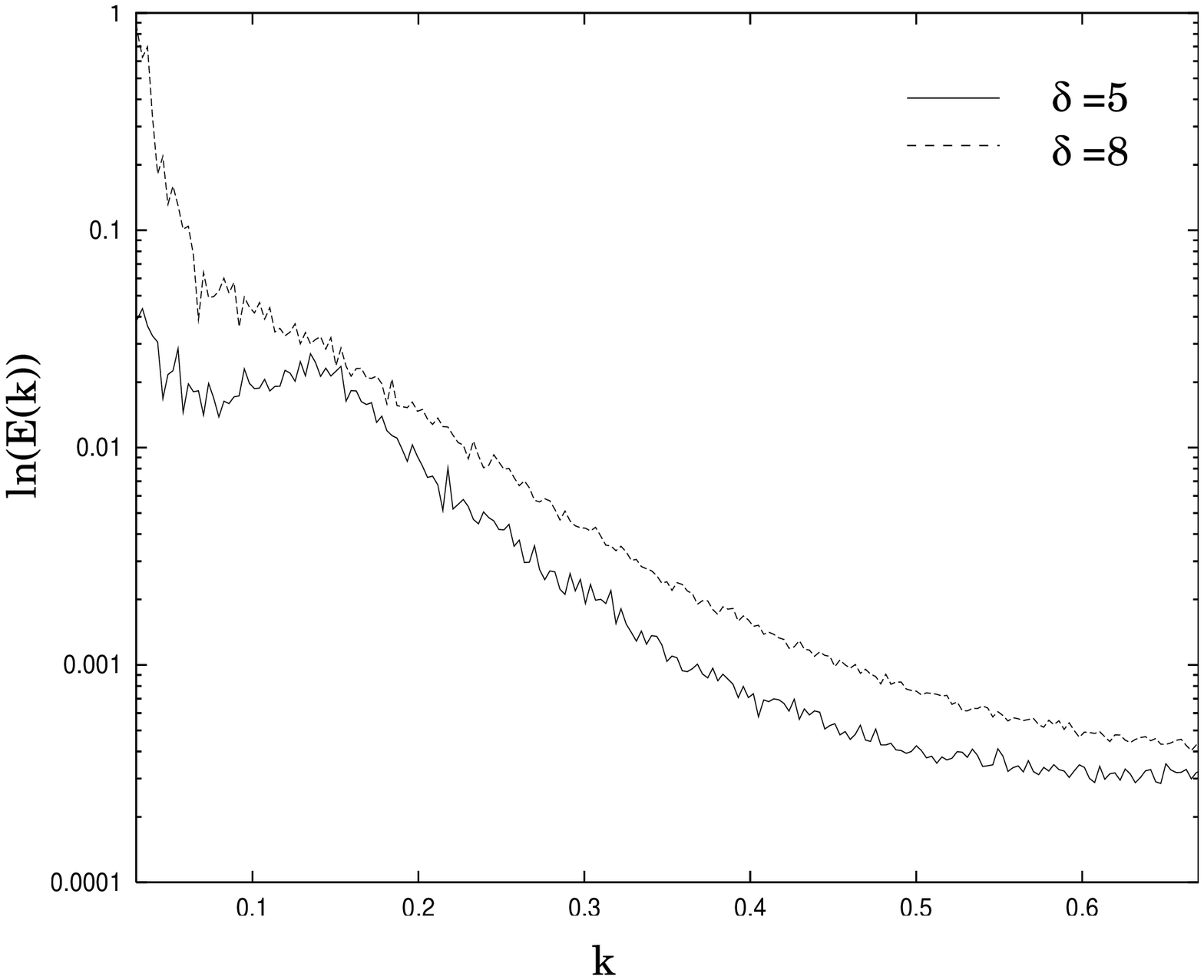}
\end{figure}
%\newpage
%\begin{figure}[h]
%\epsfxsize = \textwidth
%\epsfbox{fig4.eps}
%\end{figure}
\newpage
\begin{figure}[h]
\epsfxsize = \textwidth
\epsfbox{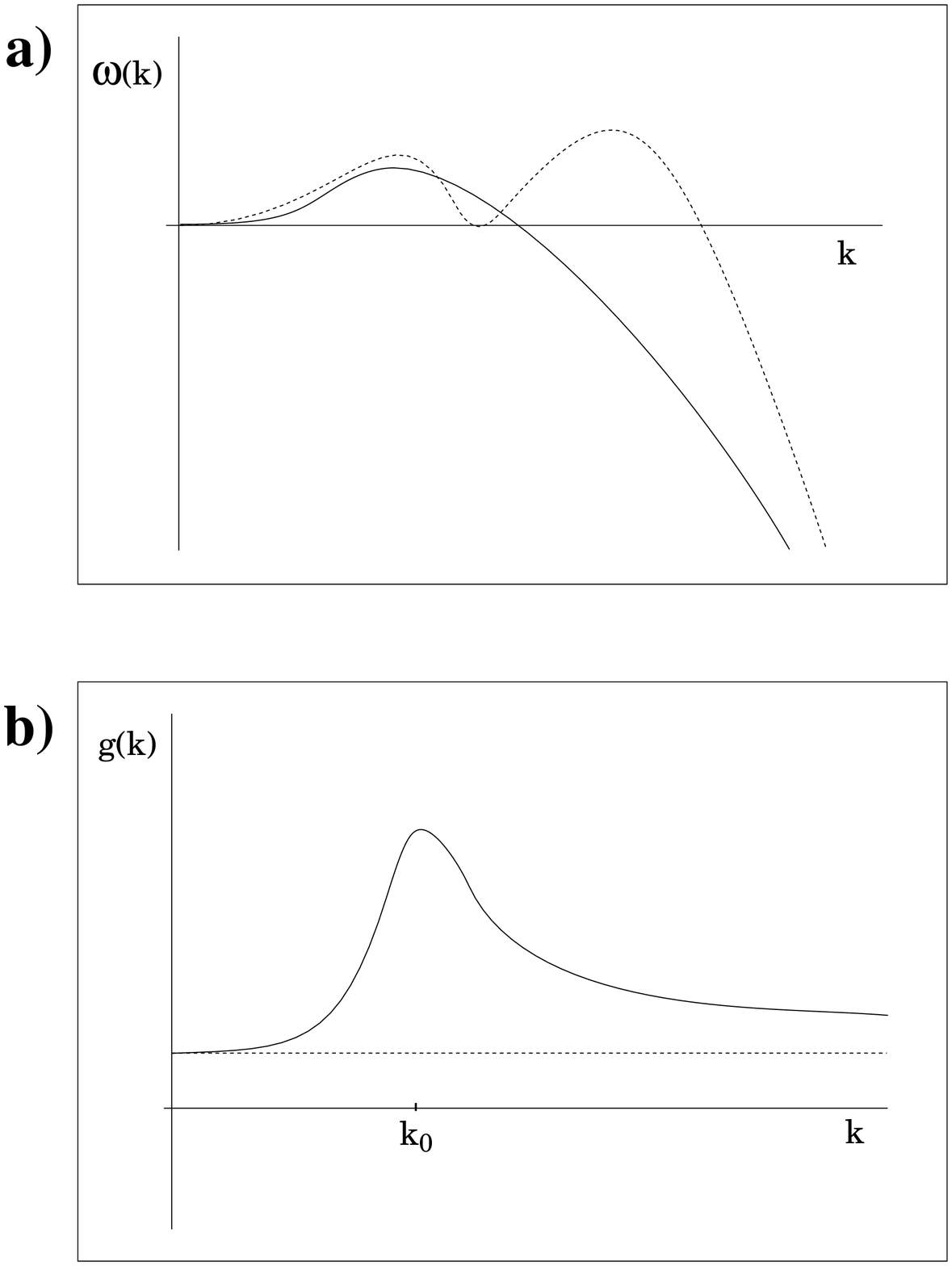}
\end{figure}
\newpage
\begin{figure}[h]
\epsfxsize = \textwidth
\epsfbox{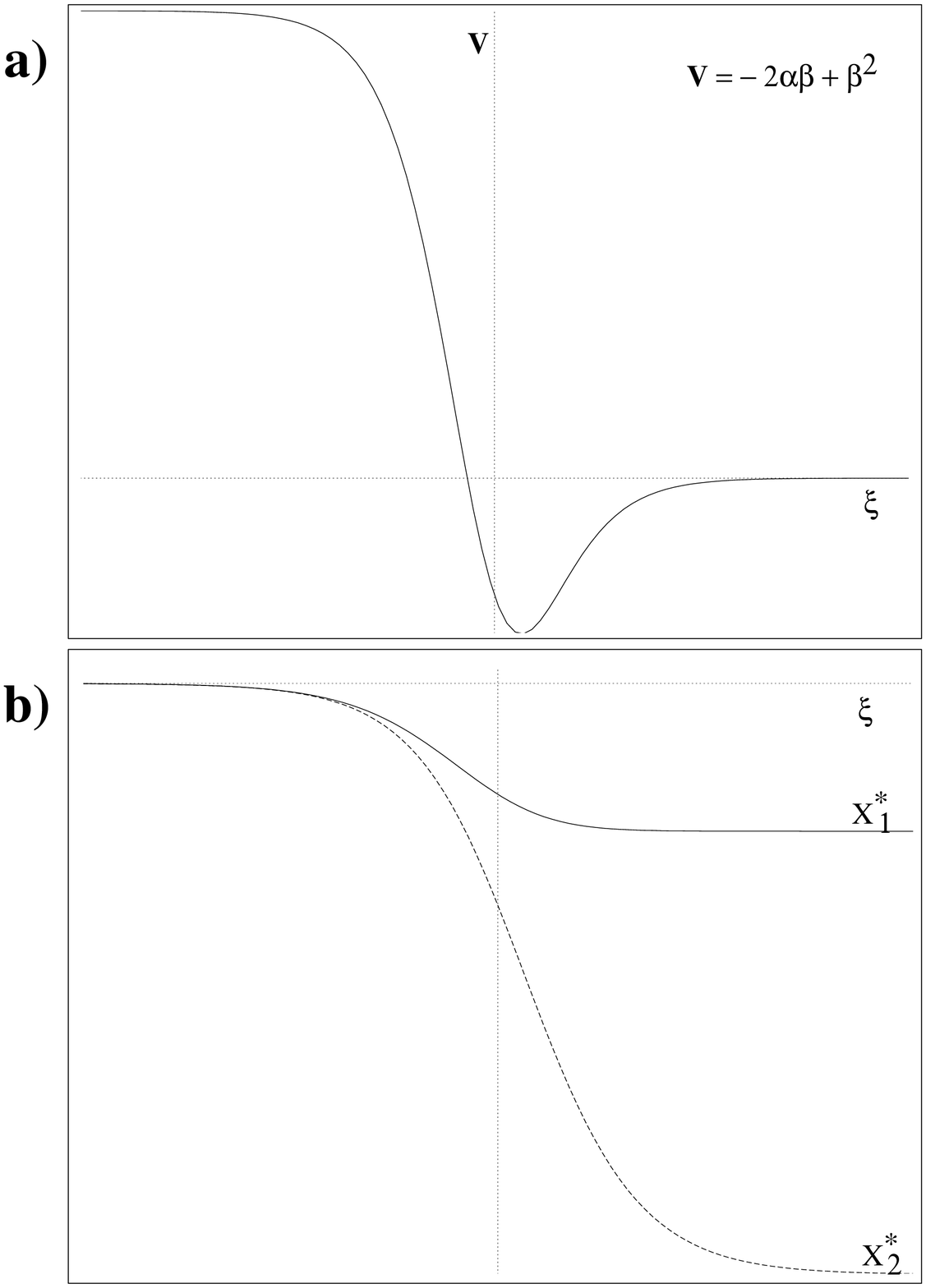}
\end{figure}
\newpage
\begin{figure}[h]
\epsfxsize = \textwidth
\epsfbox{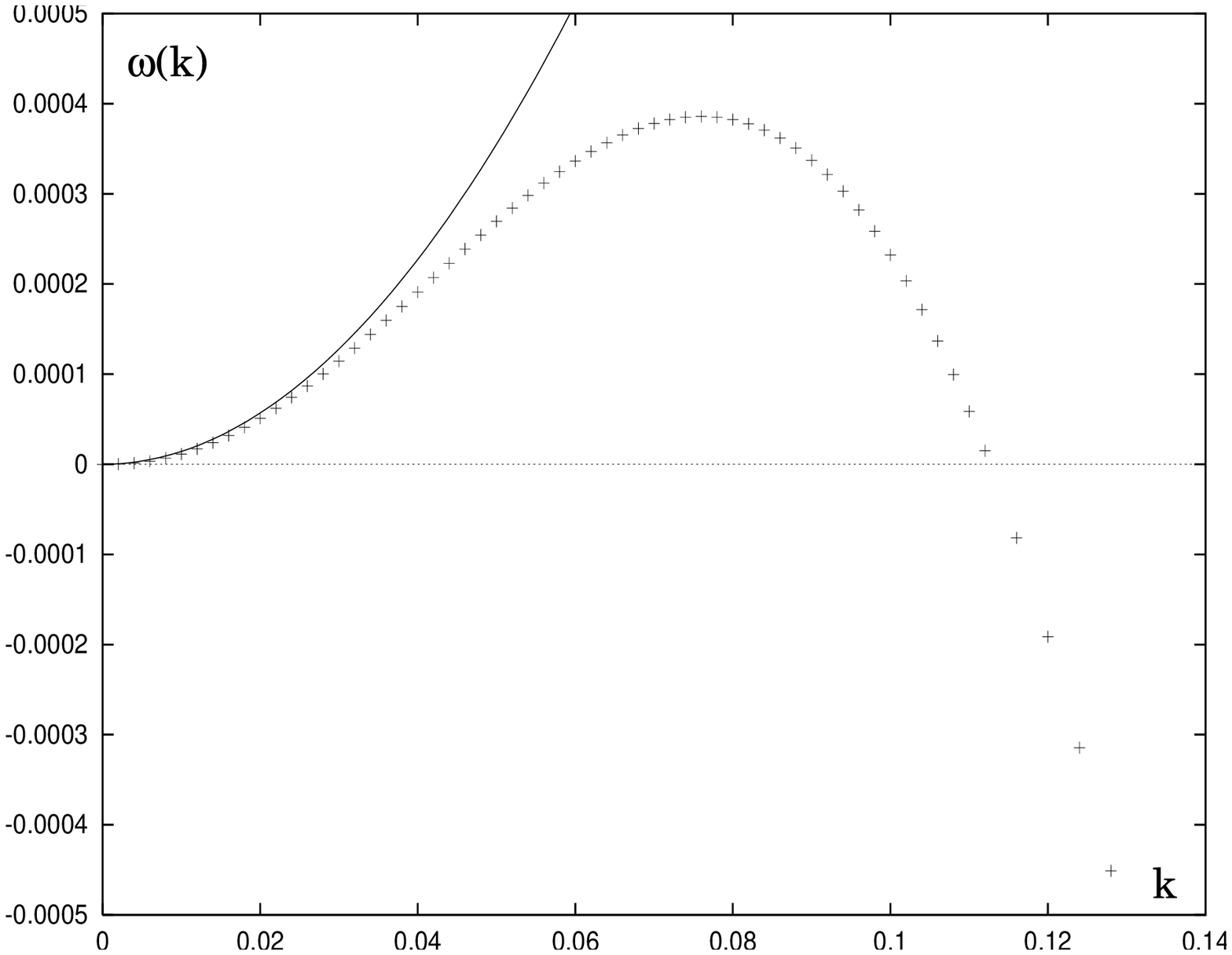}
\end{figure}


\begin{thebibliography}{1}

\bibitem[*]{by;ine} Permament address: Departament de Qu\'\i mica F\'\i sica,
 Universitat de Barcelona, Diagonal 647,  E-08028 Barcelona, Spain.

\bibitem{stan:gro} See, for instance, A.-L. Barab\'asi and H. E. Stanley, 
{\em Fractal Concepts in Surface Growth}, (Cambridge University Press, Cambridge, 
1995).

\bibitem{flames} {\em Non-Steady Flame Propagation}, ed. G. H. Markstein, 
(Macmillan, New York, 1964).

\bibitem{meron} See, for example, A. Arneodo and J. Elezgaray, 
in {\em Chemical Waves and Patterns}, eds. R. Kapral and K.
Showalter, (Kluwer, Dordrecht, 1995), p. 517; A. Hagberg and E. Meron, 
Phys. Rev. Lett. {\bf 72}, 2494 (1994).

\bibitem{siva:inst} See G. I. Sivashinsky, {\em Ann. Rev. Fluid Mech.}
{\bf 15}, 179 (1983), and references therein.

\bibitem{horv:ins2} D. Horv\'ath and K. Showalter, J. Chem. Phys. 
{\bf 102}, 2471 (1995).

\bibitem{kur:orig} Y. Kuramoto and T. Tsuzuki, Prog. Theor. Phys. {\bf 55}, 
356 (1976).

\bibitem{siv:orig} G. I. Sivashinsky, Combust. Sci. Tech. {\bf 15}, 
137 (1977).

\bibitem{bill:deve}
J. Billingham and D. J. Needham, {\em Phil. Trans. R. Soc., Ser. A} 
{\bf 334}, 1 (1991).

\bibitem{scot:simp} S. K. Scott and K. Showalter, J. Phys. Chem. {\bf 96}, 
8702 (1992).

\bibitem{horv:inst}
D. Horv\'ath, V. Petrov, S. K. Scott and K. Showalter, 
{\em J. Chem. Phys.} {\bf 98}, 6332 (1993).

\bibitem{euler} These plots were produced by an Euler integration of
(\ref{rdeq}) with space and time steps $\Delta r$ and $\Delta t$, selected 
to insure numerical stability and convergence, which was verified by 
varying these parameters. Typical values are $\Delta r=2$ and 
$\Delta t=0.5$. In order to mimic a system infinite in the $x$ direction,
the lattice was shifted to compensate for the front motion so that the
interface was always contained in the finite integration domain.

\bibitem{kura:chem} Y. Kuramoto, {\em Chemical Oscillations, Waves, 
and Turbulence.}, (Springer-Verlag, New York, 1984).

\bibitem{foot1} From the assertion about the stability of the one 
dimensional problem we have
$ \Re(\lambda_i)< \sigma < 0 $ and we see that, for small values of 
$\nu$ in the Kuramoto-Sivashinsky equation,
$ k \sim \langle{\bf u}_0|{\bf D}|{\bf u}_0\rangle^2 $ is also small
and the procedure is self-consistent in this limit.

\bibitem{pik:bohr} See also, T. Bohr and A. Pikovski, Phys. Rev. Lett. 
{\bf 70}, 2892 (1993).

\bibitem{yak:lar} V. Yakhot, Phys. Rev. A {\bf 24}, 642 (1981).

\bibitem{kpz} M. Kardar, G. Parisi and Y.-C. Zhang, Phys. Rev. A {\bf 39}, 
3053 (1989).

\bibitem{pro:red} I. Procaccia, M. H. Jensen, V. S. L'vov, K. Sneppen and
R. Zeitak, Phys. Rev. A {\bf 46}, 3220 (1992).

\bibitem{sne:dyn} K. Sneppen, J. Krug, M. H. Jensen, C. Jayaprakash and T.
Bohr, Phys. Rev. A {\bf 46}, R7351 (1992).

\bibitem{zale:stoc} S. Zaleski, {\em Physica D}, {\bf 34}, 427 (1989).

\bibitem{koga:loca} Koga and Kuramoto [Prog. Theor. Phys. {\bf 63}, 106 (1980)] 
carried out a stability analysis of the stationary,
localized solution of the FitzHugh-Nagumo model. In this case the solutions
of the adjoint problem are also localized and no difficulties are encountered.  
In the stability analysis of a propagating
front the left eigenvector does not necessarily decay as distances tend to infinity.
There is another problem arising from the chemical nature of the model:
the reaction-diffusion equation possesses an
additional zero left eigenvector due to the conservation of the reagents
and no recipe is given for how to choose the correct one.

\bibitem{foot2} 
Integration of the Bohr momentum over the allowable domain yields
$3.1$ when the next value in quantization rule is 
$\displaystyle 3\pi/2 \approx 4.5$.

\bibitem{foot3} We write $|{\bf x}\rangle$ as
$ |{\bf x}\rangle = \sum\limits_{i>0} \eta_i |{\bf u}_i\rangle $ so that
$ \left[{\bf A}|{\bf u}_0\rangle-\langle{\bf u}_0|{\bf A}|{\bf u}_0\rangle
|{\bf u}_0\rangle \right] =
\sum\limits_{i>0} \lambda_i\eta_i \cdot |{\bf u}_i\rangle $.
$ \langle{\bf u}_0|{\bf B}|{\bf x}\rangle $ is replaced by
$ \sum\limits_{i>0} \eta_i \langle{\bf u}_0|{\bf B}|{\bf u}_i\rangle $ but
$ \lambda_i\eta_i =
\langle{\bf u}_i|{\bf A}|{\bf u}_0\rangle $, thus proving the statement.

\bibitem{foota} In the simulations we used the  NAG library 
for numerical analyses.  Linear
equations were solved using the NAG library routine F04FAF.  For
interpolation we employed subroutine E01BEF which computes a
monotonicity-preserving, piecewise-cubic, Hermite interpolant to a set
of data points. The value of the definite integral of the interpolant
was determined using E01BHF.

\bibitem{zhan:stab} Z. Zhang and {S. A. E. G.} Falle, Proc.R. Soc., Ser. A 
{\bf 446}, 1 (1994).

\bibitem{milt:inst} {R. A.} Milton and {S. K.} Scott, J. Chem. Phys. 
{\bf 102}, 5271 (1995).
 
\bibitem{foot4} One may be tempted to look for stability of the system
with respect of some class of perturbations and thus avoid
computation of the left zero eigenvector. In this case the value
obtained is unpredictable. Take, for example, a perturbed solution in the
form $ {\bf z}(t,x,y) = {\bf z}_0(x-ct+\mbox{\boldmath{$\eta$}(t,y)})$ 
for the cubic autocatalysis model. Expanding
equation (\ref{rdeq}) in a power series in \mbox{\boldmath{$\eta$}} and
integrating over all $x$ we arrive at the set of equations
\begin{displaymath}
\begin{array}{rcl}
	\partial \eta_1/\partial t & = & A(\eta_1-\eta_2)+D_A\Delta \eta_1\\
	\partial \eta_2/\partial t & = & A(\eta_1-\eta_2)+D_B\Delta \eta_2\;,
\end{array}
\end{displaymath}
where $ A $ is some positive constant and
we have used $\displaystyle \int 2\alpha \beta \beta_{\xi} d\xi = -\int
\beta^2\alpha_{\xi} d\xi $.
The above system loses stability whenever $ D_A > D_B $ and it is obvious
that this value does not coincide with the critical diffusion
coefficient ratio.

\bibitem{foot6} We use the set of identities:
\begin{displaymath}
\begin{array}{rcl}
-c\alpha_{\xi} & = & (1+\mu)\alpha_{\xi\xi} - \beta^2\alpha_{\xi} -
2\alpha\beta\beta_{\xi} \\
-c\beta_{\xi} & = & (1-\mu)\beta_{\xi\xi} + \beta^2\alpha_{\xi} +
2\alpha\beta\beta_{\xi}
\end{array}
\end{displaymath}
obtained by differentiation of the front solution.  We find the system
(\ref{disp}) to be more compact when we use this particular form of the 
perturbation.

\end{thebibliography}
\end{document}